\documentclass[11pt]{article}

\usepackage{mathrsfs,amsmath,amsbsy,amssymb,mcite}

\def\be{\begin{equation}}
\def\ee{\end{equation}}

\def\nn{\nonumber \\}
\def\na{\nabla}
\def\mc{\mathcal}
\def\dag{\dagger}

\DeclareTextFontCommand{\textwasy}{\wasyfamily}
\def \wasyfamily{\fontencoding{U}\fontfamily{wasy}\selectfont}
\def \thorn{{\wasyfamily\char105}}
\DeclareTextCommand{\dh}{OT1}{{\wasyfamily\char107}}
\newcommand{\tho}{{\textrm\thorn}}
\renewcommand{\eth}{{\textrm{\dh}}}
\newcommand{\thop}{\tho'}

\newcommand{\half}{{\textstyle{\frac{1}{2}}}}
\renewcommand{\nn}{{\nonumber}}

\newcommand{\del}{{\delta}}

\newcommand{\kap}{{\kappa}}

\newcommand{\La}{{\Lambda}}

\newcommand{\Om}{\Omega} 
\newcommand{\Ps}{\Psi}   
\newcommand{\vphi}{\varphi} 
\newcommand{\Phis}{\Phi^\mathrm{S}} 
\newcommand{\Phia}{\Phi^\mathrm{A}} 

\newcommand{\lb}{{\ell}}
\newcommand{\nb}{{n}}
\newcommand{\mb}[1]{{m_{(#1)}}}

\newcommand{\M}[1]{{\stackrel{#1}{M}}}  

\numberwithin{equation}{section}

\title{The perturbation theory of higher dimensional spacetimes\\
 \textit{\'{a} la} Teukolsky} 

\author{Mahdi Godazgar \\Department of Applied Mathematics and Theoretical Physics \\ Centre for Mathematical Sciences \\ Wilberforce Road, Cambridge CB3 0WA, UK\\ mmg31@cam.ac.uk}

\begin{document}

\maketitle

\begin{abstract}
We consider the possibility of deriving a decoupled equation in terms of Weyl tensor components for gravitational perturbations of the 
Schwarzschild-Tangherlini solution. We find a particular gauge invariant component of the Weyl tensor 
does decouple and argue that this corresponds to the vector modes of Ishibashi and Kodama.  
Also, we construct a Hertz potential map for solutions of the electromagnetic and gravitational perturbation 
equations of a higher dimensional Kundt background using the decoupled equation of Durkee and Reall. 
Motivated by recent work of Guica and Strominger, we use this to construct the asymptotic 
behaviour of metric perturbations of the near-horizon geometry of the 5d cohomogeneity-1 Myers-Perry black hole.
\end{abstract}

\section{Introduction}

Soon after the discovery, by Kerr, of a solution to the vacuum Einstein equation representing an isolated rotating black hole \cite{kerr}, the status of its classical stability spawned an area of active research in general relativity. Significant progress in this direction was made by Teukolsky, who realised that in the case of algebraically special solutions \cite{cartan,*ruse,*pet,*debever,*pen,*pirani, steph}, of which the Kerr solution is an example, one is able to derive a decoupled equation, satisfied by Weyl scalar $\Psi_0$, from the original perturbation equation \cite{teuk, teuk1}. The validity of a separability ansatz, which is related to the existence of hidden symmetries \cite{carter, walkpen}, allowed Press and Teukolsky to use the decoupled equation to provide strong evidence for the linear (mode) stability of the Kerr solution under non-algebraically special perturbations \cite{teuk2}.

The existence of a decoupled equation was later shown to be related to the existence of gauge invariant quantities \cite{stewalk}. Given that the Weyl scalar $\Psi_0$, which solves the decoupled equation, is gauge invariant and has the same number of degrees of freedom as the perturbed metric, \textit{viz.} 2, one may suspect that $\Psi_0$ encodes all information regarding the perturbation, i.e. a solution of the perturbation equation can be constructed given the existence of a decoupled equation.  This was shown to be true by Kegeles and Cohen \cite{kc, kc2} and Chrzanowski \cite{chr}.  They outlined a constructive procedure (the Hertz map) for finding solutions of perturbation equations for an algebraically special background given some technical assumptions.  A very short and elegant proof of these statements was provided by Wald some time later \cite{wald}.

The study of higher dimensional gravity, particularly its black hole solutions, has become an active area of research in recent years.  One of the most natural questions that one can ask is whether known black holes solutions, such as the Myers-Perry \cite{mp}, or the black ring \cite{bring} solutions are classically stable.  The stability of the Schwarzschild-Tangherlini solution, which can be thought of as belonging to the Myers-Perry family has been demonstrated \cite{ishi}.

However, it has been known for some time, by analogy with Gregory-Laflamme (G-L) type instabilities \cite{gl}, that one would expect instabilities to occur in certain other regimes of the Myers-Perry and black ring families.  For example, one would expect thin black rings to suffer from the same type of G-L instability that occurs for a black string \cite{bring}.  Or, for ultraspinning ($d \geq 6$) Myers-Perry black holes to suffer from the same kind of G-L type instability that one finds for $p$-branes \cite{empmyers03}. There has been a great deal of recent progress in using numerical methods to tackle and confirm these conjectures, at least, as far as the Myers-Perry solution is concerned \cite{dias09, *shibata, *dias10, *shibata2, *dias210}.

Nevertheless, experience from 4d GR suggests that there may be a more simple framework (\textit{\`{a} la} Teukolsky) in which the stability of higher dimensional black hole solutions can be addressed.  The motivation of such a consideration would not only be that such a framework would facilitate a much simpler study of perturbations of regimes currently under investigation using numerical methods but that it may allow a study of regimes that are currently inaccessible to numerical investigations. 

The higher dimensional generalisation of the Teukolsky decoupled equation was first considered in Ref. \cite{hdecoup}.  In higher dimensions, one can construct the analogue of $\Psi_0$ by choosing a null frame with null vectors
$\ell$ and $n$ such that $\ell \cdot n =1$ and complete this frame with $d-2$ orthonormal spacelike vectors $\mb i \ (i=2,\ldots,d-1)$ that are orthogonal to $\ell$ and $n$.  Now, we define the higher dimensional
generalisation of $\Psi_0$ to be $\Om_{ij}=\ell^a \mb i^b \ell^c \mb j^d C_{abcd}$\footnote{$\Omega_{ij}$ is the higher dimensional
 generalisation of $\Psi_{0}$ in the sense that they are both the boost weight +2 components of the Weyl tensor.}.  Furthermore, for algebraically special
solutions the perturbed value of $\Om_{ij}$ is gauge invariant (under infinitesimal diffeomorphisms and frame transformations) \cite{hdecoup}.  Thus, as the higher dimensional 
generalisation of $\Psi_0$, $\Omega_{ij}$ is the most natural candidate to consider decoupling in higher dimensions.

By studying the conditions required for $\Om_{ij}$ to decouple, it was found that in higher dimensions, it is not sufficient for the background solution to be algebraically special.  In addition, there must exist a geodesic null congruence with vanishing expansion, shear and twist.  Equivalently, the solution must be Kundt. 

With the above motivation in mind, the strong restrictions found in Ref. \cite{hdecoup}, which a background must satisfy for $\Om_{ij}$ to decouple, 
can be thought of as being disappointing.  However, this result can still be used to study perturbations of a particular class of 
higher dimensional black holes.  This is because the near-horizon geometry (NHG) of extreme black hole solutions are known to be Kundt.  This fact was used by Durkee and Reall to find instabilities of the NHG of cohomogeneity-1 Myers-Perry solutions \cite{durkee10}.  
A conjecture relating instabilities of the NHG and the full solution for perturbations preserving certain symmetries was then used to predict an instability of extremal and near-extremal cohomogeneity-1 Myers-Perry solutions in seven or more dimensions \cite{durkee10}.  These predictions were later confirmed in Ref. \cite{dias11}.

Even so, one would like a decoupled equation that allows the study of a greater class of solutions within the family of the Myers-Perry solution.  While, it is true that $\Omega_{ij}$ does not decouple it may be possible to construct some other gauge invariant quantity from the Weyl tensor that does. The first aim of this paper is to consider this possibility for the Schwarzschild-Tangherlini background as a simple class within the Myers-Perry family.  The existence of gauge invariant quantities constructed from the Weyl tensor that decouple on the Schwarzschild-Tangherlini background would give us hope of finding analogous quantities for more general solutions within the Myers-Perry family.  Equivalently, the absence of such quantities for the Schwarzschild-Tangherlini solution would indicate their absence for more general solutions. 

In section \ref{decoup}, we begin by considering whether other gauge invariant quantities can be added to $\Om_{ij}$, so that the new quantity decouples.  The strategy we take is to find the obstruction to the decoupling of $\Om_{ij}$. Then, we consider the decoupling of this new quantity and find the obstruction to its decoupling.  Once this iterative process terminates we are left with a set of gauge invariant quantities that form an obstruction to each others' decoupling. For the case of $\Om_{ij}$, there are three basic gauge invariant quantities including $\Om_{ij}$ that must be considered. However, we find that a linear combination of these quantities can never decouple.  Hence, there is no hope of constructing a gauge invariant quantity that decouples using $\Om_{ij}$. 

We then consider other gauge invariant quantities constructed from components of the Weyl tensor and find that a particular set of components $\Phia_{ij} \equiv \ell^a n^b \mb{[i}^c \mb{j]}^d C_{abcd}$ do decouple. Metric perturbations of the Schwarzschild-Tangherlini solution have been studied by Ishibashi and Kodama \cite{ishi} using a gauge invariant analysis that is analogous to the 4d gauge invariant approach developed by Moncrief \cite{mon}. This approach uses the spherical symmetry of the solution to construct gauge invariant quantities from the metric perturbations. These quantities can be classified into tensor, vector and scalar modes depending on how they behave on the sphere.  We argue that $\Phia_{ij}$ is related to the vector modes of Ishibashi and Kodama.  

This is completely analogous to what is known to happen in 4d. On the Schwarzschild background, the perturbed value of the imaginary part of the Weyl scalar $\Psi_2$, which is gauge invariant and has $\Phia_{ij}$ as its higher dimensional generalisation, satisfies a decoupled equation. Furthermore, this decoupled equation is equivalent to the Regge-Wheeler equation \cite{regge} describing vector mode (or axial) perturbations of the Schwarzschild background \cite{price}.

Our results decrease the likelihood of finding a gauge invariant quantity constructed from Weyl tensor components that decouples on the more general Myers-Perry background.  This is because $\Phia_{ij}$ is generally non-zero for the Myers-Perry solution, so its perturbation is not gauge invariant.  It is not inconceivable that there exists a gauge invariant quantity constructed from $\Phia_{ij}$ that decouples.  However, we have not been able to come up with a suitable candidate.

The second aim of the paper is to construct the Hertz potential map for constructing solutions of the perturbation equations of higher dimensional Kundt backgrounds in the manner proposed by Wald \cite{wald}.

In section \ref{hertz}, we begin by presenting Wald's argument for constructing solutions of a general perturbation equation given a decoupled equation.  It was shown in \cite{hdecoup} that a decoupled equation for electromagnetic and gravitational perturbations exist for Kundt backgrounds. Thus, we apply this method to construct the Hertz map for electromagnetic  (section \ref{elec}) and gravitational (section \ref{grav}) perturbations of Kundt backgrounds in higher dimensions.

As stated above, the NHG of extremal black hole solutions is Kundt.  In particular, the decoupled equation for NHG of cohomogeneity-1 Myers-Perry solutions was studied in \cite{durkee10}.  As an application, we use the results of \cite{durkee10} to determine the asymptotic behaviour of metric perturbations of the NHG of the 5d cohomogeneity-1 Myers-Perry black hole.  This is, in part, motivated by a recent paper concerning the entropy counting of such black hole solutions \cite{guica5d}.  We find that there exist modes that violate the boundary conditions required in Ref. \cite{guica5d}.  Thus, at higher orders, it will not be possible to deform the NHG.

\section{Decoupling results for perturbations of the Schwarzschild-\\Tangherlini solution} \label{decoup}

In this paper, we define the gauge invariance of a perturbed quantity following Stewart and Walker \cite{stewalk}.  Given a spacetime 
scalar $X$, we can decompose this into its background and perturbed value, i.e. to first order $X=X^{(0)} + X^{(1)}$.  We are interested in the how 
the perturbed value of $X$, $X^{(1)}$ changes under infinitesimal coordinate and frame transformations.  We say that $X^{(1)}$ is gauge invariant if and only if it 
remains invariant under coordinate and frame transformations. This condition places constraints on the background value of $X$.  For example, under a coordinate 
transformation parametrised by $\xi$, $X^{(1)} \rightarrow X^{(1)} + \xi \cdot \partial X^{(0)}$. Hence, for $X^{(1)}$ to remain invariant under infinitesimal coordinate transformations 
$X^{(0)}$ must be constant.  The invariance of the perturbed quantity $X^{(1)}$ under infinitesimal frame transformations is verified by performing infinitesimal frame transformations given in equations \eqref{app:boost}---\eqref{app:rotn} of appendix \ref{app:ghp}. In Ref. \cite{hdecoup}, it was shown that  the perturbed value of $\Om_{ij}$ is gauge invariant if and only if the background solution is algebraically special.  Similar arguments to those presented in Ref. \cite{hdecoup} can be used to demonstrate the gauge invariance of other Weyl components given a background.  For brevity, we will suppress superscript labels on quantities indicating whether they are background or perturbed values since this will be apparent from the context.

In Ref. \cite{hdecoup}, Durkee and Reall consider the status of Teukolsky's decoupling result in higher dimensions.  They do this by taking an algebraically special background, in which the perturbed value of $\Om_{ij}$ is gauge invariant. Using the Bianchi identity equations of the higher dimensional GHP formalism\footnote{See appendix \ref{app:ghp} for a review of the higher dimensional GHP formalism.} \cite{ghp}, they derive a second order coupled differential equation, where all second order derivatives act on the gauge invariant quantity $\Om_{ij}$.   For $\Om_{ij}$ to decouple, all other terms involving other components of the Weyl tensor must either be re-expressed in terms of terms involving $\Om_{ij}$ or vanish.  Once all the Newman-Penrose and Bianchi equations have been exhausted, they conclude that for $\Om_{ij}$ to decouple, the background must be Kundt.  This means that the background admits a null geodesic congruence with vanishing expansion, shear and twist.

Here, we shall take a different approach.  Instead of forcing a particular gauge invariant quantity (in their case $\Om_{ij}$) to decouple and finding restrictions on the background, we fix the background (in this case the Schwarzschild-Tangherlini solution) and try to find gauge invariant quantities constructed from Weyl tensor components that decouple on this background.

Thus, we will use the higher dimensional GHP formalism developed in Ref. \cite{ghp} to address the possibility that there exists a gauge invariant quantity constructed from Weyl tensor components that decouples on the Schwarzschild-Tangherlini background \footnote{See appendix \ref{app:schw} for a review of the Schwarzschild-Tangherlini solution.}.

We begin by considering the derivation of the decoupling result in \cite{hdecoup} for the Schwarzschild-Tangherlini 
background.  Up to equation (5.31) in \cite{hdecoup}, the only assumption made is that the background admits a null geodesic congruence with vanishing shear and rotation. This is satisfied by the Schwarzschild-Tangherlini solution.  Removing other second order terms from equation (5.31) gives
\be \label{decoupOm}
\left(2\tho'\tho+ \eth_k \eth_k + \rho'\tho + {\textstyle \frac{d+6}{d-2}} \rho \tho' + {\textstyle \frac{2}{d-2}} \rho \rho' + {\textstyle \frac{4(d-2)}{d-3}} \Phi \right) \Om_{ij} 
+ { \textstyle \frac{2 \rho^2}{d-2} } \left( \Phis_{ij}- { \textstyle \frac{\Phi }{d-2}} \delta_{ij} \right) =0.
\ee
Thus, the trace-free part of $\Phis_{ij}$ is the obstruction for the decoupling of $\Omega_{ij}$ \footnote{Note that in 4d this term vanishes identically leading to the decoupling of $\Psi_0$.}. We know that this vanishes on the background (see appendix \ref{app:schw}).  Hence, it is invariant under coordinate transformations.  By applying infinitesimal frame transformations \eqref{app:boost}---\eqref{app:rotn} and using the fact that the background solution is type D, we can simply verify that $\Phis_{ij}- { \textstyle \frac{\Phi }{d-2}} \delta_{ij}$ is a gauge invariant quantity.

We proceed by finding the obstruction to the decoupling of the gauge invariant quantity $$ \rho^2 \left( \Phis_{ij}- { \textstyle \frac{\Phi }{d-2}} \delta_{ij} \right). $$
It can be shown that this quantity satisfies\footnote{The derivation of this equation is given in appendix \ref{app:deriv}.}
\begin{align}
\left(2\tho'\tho+ \eth_k \eth_k + {\textstyle \frac{d+6}{d-2}} (\rho' \tho + \rho \thop) + {\textstyle \frac{4 \Phi}{\rho}} \tho + {\textstyle \frac{2(3d+5)}{(d-2)^2}} \rho \rho' - {\textstyle \frac{4(3d-8)}{(d-2)(d-3)}} \Phi \right) & \left[ \rho^2 \left(\Phis_{ij} - {\textstyle \frac{\Phi}{d-2}} \delta_{ij} \right) \right]  \nn \\ \label{rphitdecoup}
 & \hspace{-10mm} + {\textstyle \frac{(d-4)}{(d-2)^2}} \rho^2 \left( \rho'^2 \Om_{ij}+ \rho^2 \Om'_{ij} \right) =0.
\end{align}
Hence, the obstructions to the decoupling of $\rho^2 \left(\Phis_{ij} - {\textstyle \frac{\Phi}{d-2}} \delta_{ij} \right)$ are $\rho^2 \rho'^2 \Om_{ij}$ and $\rho^4 \Om'_{ij}$, which can simply be shown to be gauge invariant.  An equation of the form above can be derived for these quantities using equation \eqref{decoupOm}.  However, this procedure will lead to yet another new quantity to consider: $\rho^4 \rho'^2 \left(\Phis_{ij} - {\textstyle \frac{\Phi}{d-2}} \delta_{ij} \right)$.  

Thus, we find that this iterative process generates boost weight +2 terms of the form $\Om_{ij}$, $\rho^2 \left(\Phis_{ij} - {\textstyle \frac{\Phi}{d-2}} \delta_{ij} \right)$ and $\rho^4 \Om'_{ij}$ multiplied by various factors of $\rho \rho'$.  Therefore, we take 
\be \label{decoupans}
X_{ij} = \Om_{ij} + f \ \rho^2 \left(\Phis_{ij} - {\textstyle \frac{\Phi}{d-2}} \delta_{ij} \right) + g \ \rho^4 \Om'_{ij}
\ee
as our ansatz for the decoupled gauge invariant quantity, where $f$ and $g$ are GHP scalars of boost and spin weight 0.  

For $X_{ij}$ to decouple, there must exist $f$ and $g$ such that
\be
\left(2 \thop \tho + \eth_{k} \eth_{k} \right)X_{ij} = \mc O X_{ij},
\ee
where $\mc O$ is some first order differential operator.

The only boost and spin weight 0 GHP scalars that do not vanish on the background are $\varsigma=\rho \rho'$ and $\Phi$.  Hence,
\be
f=f(\varsigma,\Phi), \quad g=g(\varsigma,\Phi).
\ee
In particular, equations \eqref{NP3}, \eqref{NP3}$'$ and \eqref{B5} evaluated on the background imply that on the background
\be
\eth_{i} f = 0, \quad \eth_{i} g = 0.
\ee

Using equation \eqref{decoupOm}, its primed version for $\rho^4 \Om'_{ij}$ and eq. \eqref{rphitdecoup}, we find
\begin{align}
 \left(2 \thop \tho + \eth_{k} \eth_{k} \right)X_{ij} = & \left\lbrace \rho' \tho + \textstyle{\frac{d+6}{d-2}} \thop + \ldots \right\rbrace \Om_{ij} \nn \\
+& \left\lbrace \left(-2 \textstyle{\frac{\thop f}{f}} + \textstyle{\frac{d+6}{d-2}} \rho' + \textstyle{\frac{4\Phi}{\rho}} \right) \tho + \left( -2 \textstyle{\frac{\tho f}{f}} + \textstyle{\frac{d+6}{d-2}} \rho  \right) \thop + \ldots \right\rbrace \left[ f \rho^2 \left(\Phis_{ij} - {\textstyle \frac{\Phi}{d-2}} \delta_{ij} \right) \right] \nn \\
+& \left\lbrace \left(-2 \textstyle{\frac{\thop g}{g}} + \textstyle{\frac{d+14}{d-2}} \rho' + \textstyle{\frac{8\Phi}{\rho}} \right) \tho + \left( -2 \textstyle{\frac{\tho g}{g}} + \textstyle{\frac{d+6}{d-2}} \rho  \right) \thop + \ldots \right\rbrace \left( g \rho^4 \Om'_{ij} \right),
\end{align}
where the ellipses indicate other scalar terms that have been omitted for brevity.  Necessarily, for $X_{ij}$ to decouple, the coefficients of the derivative operators in the 3 terms on the right hand side must be equal, i.e.
\begin{gather}
 \rho' = -2 \textstyle{\frac{\thop f}{f}} + \textstyle{\frac{d+6}{d-2}} \rho' + \textstyle{\frac{4\Phi}{\rho}}  = -2 \textstyle{\frac{\thop g}{g}} + \textstyle{\frac{d+14}{d-2}} \rho' + \textstyle{\frac{8\Phi}{\rho}}, \nn \\
\textstyle{\frac{d+6}{d-2}} \rho = -2 \textstyle{\frac{\tho f}{f}} + \textstyle{\frac{d+6}{d-2}} \rho = -2 \textstyle{\frac{\tho g}{g}} + \textstyle{\frac{d+6}{d-2}} \rho.
\end{gather}
These imply that on the background
\be
\tho f=0, \quad \tho g =0, \quad \thop f= 2\left( \textstyle{\frac{2}{d-2}} \rho' + \textstyle{\frac{\Phi}{\rho}} \right) f, \quad \thop g = 4\left( \textstyle{\frac{2}{d-2}} \rho' + \textstyle{\frac{\Phi}{\rho}} \right) g.
\ee 
Equation \eqref{B2} evaluated on the background implies that
\be
\thop \Phi  =- \textstyle{\frac{d-1}{d-2}} \rho \Phi.
\ee 
This, along with the equations above, can be used to show that
\be
[\thop,\tho]f=\left( \textstyle{\frac{4}{(d-2)^2}}\rho \rho' + \textstyle{\frac{2(d-1)}{d-2}} \Phi - \textstyle{\frac{1}{d-2}}\right) f,
\ee
which is non-vanishing on the background.  However, considering commutator equation \eqref{C1} on $f$, on the background, gives that this should vanish.  Thus, we encounter a contradiction. We would arrive also at a contradiction had we considered a commutator on $g$.

Therefore, there do not exist scalars $f$ and $g$ such that the gauge invariant quantity $X_{ij}$, given by eq. \eqref{decoupans}, decouples.

In summary, we took $\Om_{ij}$ as the initial ansatz for a decoupled equation.  This does not work as there is another gauge invariant quantity that obstructs the decoupling of $\Om_{ij}$.  Then, we considered the decoupling of this new quantity and found other quantities that obstruct its decoupling. Repeating this iterative process we found that all obstructions are of a form given by one of three basic gauge invariant quantities constructed from $\Om_{ij}$, $\Phis_{ij}$ and $\Om'_{ij}$.  Thus, we take a gauge invariant quantity constructed from a linear combination of these three basic quantities as our new ansatz and find that such a quantity cannot decouple. Hence, a decoupled equation for the Schwarzschild-Tangherlini solution cannot be found by completing $\Om_{ij}$ with other gauge invariant quantities.

It is clear from the above calculation that gauge invariant quantities formed from $\Om_{ij}$ or the trace-free part of $\Phis_{ij}$ will not decouple.  However, one can consider other gauge invariant quantities.  For example, one can consider boost weight +1 components of the Weyl tensor: $\Psi_{ijk}$.  Although $\Psi_{ijk}$ is gauge invariant under infinitesimal coordinate transformations, since it vanishes on the background, its gauge invariance under infinitesimal tetrad transformations is not so clear.  Under a null rotation about $n$ parametrised by $z_{i}$, $\Psi_i$ and $\Psi_{ijk}$ transform as \cite{ghp}
\be
\Psi_{i} \mapsto \Psi_{i} - z_{i} \Phi + \ldots, \quad \Psi_{ijk} \mapsto \Psi_{ijk} + z_{[k} \Phi_{j]i} + z_l \Phi_{lijk} + \ldots,
\ee
where the ellipses refer to terms that vanish on the background and need not be considered.  Thus, in order to be gauge invariant, the boost weight +1 ansatz must be of the form
\be
\Psi_{ijk} + {\textstyle \frac{2}{d-3}} \delta_{i[j}\Psi_{k]},
\ee
which vanishes in 4d. Looking at the Bianchi equations \eqref{B1}--\eqref{B7}, it is not clear how they can be manipulated to give a decoupling result for the quantity above. 

Alternatively, we can use the Penrose wave equation \cite{penwave}
\be \label{penw}
g^{ef} \na_e \na_f R_{abcd} + R_{abef} R_{cd}^{ef}+2(R_{aecf} {R_b^e}_d^f - R_{aedf} {R_b^e}_c^f)=0
\ee 
to confirm that decoupling is indeed not possible.  Shortly after the decoupling result of Teukolsky, it was shown by Ryan, that the Teukolsky decoupled equation can be derived by projecting the Penrose wave equation on the appropriate null tetrad component and linearising \cite{ryan}.  Note, that since we are going to consider this equation to first order in perturbed quantities and, furthermore, that the perturbed solution continues to satisfy the vacuum Einstein equation, we can replace the Riemann tensors in the equation above with Weyl tensors.

Contracting the Penrose wave equation \eqref{penw} with $\ell^a \mb i^b \mb j^c \mb k^d + \textstyle{\frac{2}{d-3}} \ell^a n^b \ell^c \delta_{i[j} \mb {k]}^d$ gives
\be
\mc O \left( \Psi_{ijk} + {\textstyle \frac{2}{d-3}} \delta_{i[j}\Psi_{k]} \right) = \textstyle{\frac{2 \rho}{d-2}} \left(\eth_{i} \Phia_{jk} - \textstyle{\frac{1}{d-3}} \delta_{i[j}\eth_{k]} \Phi \right) + \ldots,
\ee
where we have used the Bianchi identities to simplify the expression on the right hand side.  The ellipses indicate terms that are of zero order in derivatives.  The Bianchi identities can not be used to transform the first order in derivatives obstruction on the right hand side into a terms involving $\Psi_{ijk} + {\textstyle \frac{2}{d-3}} \delta_{i[j}\Psi_{k]}$.  Thus, we conclude that this boost weight +1 gauge invariant quantity cannot decouple.

We will not consider gauge invariant quantities constructed from $\Phi_{ijkl}$ since it is not clear whether such a quantity would be simpler to work with than the original metric perturbation.

The only components left to consider are $\Phia_{ij}$. It is simple to verify that $\Phia_{ij}$ is gauge invariant.   We proceed by using the Bianchi equations to derive an equation, in which second order derivatives act only on $\Phia_{ij}$.  As before, we throw away terms that are clearly of quadratic order or above in the perturbation expansion.

Antisymmetrising over $ij$ in \eqref{B2} gives
\be \label{b2ant}
-\tho \Phia_{ij} + \eth_{[i} \Psi_{j]} = {\textstyle \frac{3}{d-2}} \rho \Phia_{ij} + {\textstyle \frac{d-1}{d-2}} \Phi \rho_{[ij]},
\ee
while antisymmetrising over $ij$ in \eqref{B5} gives
\be \label{b5ant}
\eth_{k} \Phia_{ij} + \eth_{[i}\Phi_{j]k} + \thop \Psi_{[ij]k}=
-{\textstyle \frac{2}{d-2}} \rho' \Psi_{[ij]k} - {\textstyle \frac{d-1}{(d-2)(d-3)}} \Phi \delta_{k[i} \tau_{j]}
-{\textstyle \frac{\rho}{d-2}} \left(\Psi'_{[ij]k}-\delta_{k[i}\Psi'_{j]} \right).
\ee
Contracting over indices $jl$ in \eqref{B3} and antisymmetrising over the remaining indices gives
\be \label{antconb3}
\eth_{[i} \Psi_{j]} - \eth_{k} \Psi_{[ij]k} = {\textstyle \frac{(d-1)(d-4)}{(d-2)(d-3)}} \Phi \rho_{[ij]}
-{\textstyle \frac{d-4}{d-2}} \rho \Phia_{ij}.
\ee
Now, consider $\eth_{k}$\eqref{b5ant} $-2\thop$\eqref{b2ant}
\be
(2\thop \tho + \eth_k \eth_k)\Phia_{ij} + \thop \eth_k \Psi_{[ij]k} -2\thop \eth_{[i}\Psi_{j]} +\eth_{[i}\eth_{|k|}\Phi_{j]k} - [\thop,\eth_k] \Psi_{[ij]k} + [\eth_k,\eth_{[i}] \Phi_{j]k} = \ldots,
\ee
where the ellipses refer to terms on the right hand side.  All such terms have only single derivatives. In addition, commutator equations \eqref{C2} and \eqref{C3} can be used to rewrite the final two terms on the right hand side as an expression involving only single derivatives.  Furthermore, equations \eqref{b5b7} and \eqref{antconb3} imply that
\[
 \thop \eth_k \Psi_{[ij]k} -2\thop \eth_{[i}\Psi_{j]} +\eth_{[i}\eth_{|k|}\Phi_{j]k} = - [\thop, \eth_{[i}] \Psi_{j]} + \ldots,
\]
which can be converted into an expression with single derivatives using \eqref{C2}$'$.  Thus, we have an equation in which all double derivatives act on $\Phia_{ij}$
\[
 (2\thop \tho + \eth_k \eth_k)\Phia_{ij} = \ldots.
\]
On the right hand side, terms of the form $\eth_{k} \Psi_{[ij]k}$ can be removed by using equation \eqref{antconb3}.  Other terms can be simplified by rearranging terms such that one of the factors in a term vanishes on the background. Then, the other factor can be assumed to take its background value.  Bianchi and Newman-Penrose equations on the background can be used to further simply such terms.

Having simplified terms on the right hand side of the equation above, we find that the expression simplifies greatly and $\Phia_{ij}$ decouples
\be
\left( 2 \thop \tho + \eth_k \eth_k + {\textstyle \frac{d+2}{d-2} } (\rho \thop + \rho' \tho) 
+ {\textstyle \frac{4(d-1)}{(d-2)^2} } \rho \rho' - {\textstyle \frac{2(d-1)}{d-3} } \Phi \right) \Phia_{ij} =0 .
\ee 

Metric perturbations of the Schwarzschild-Tangherlini solution have been studied by Ishibashi and Kodama \cite{ishi} using a gauge invariant analysis that is analogous to the 4d approach developed by Moncrief \cite{mon}. $\Phia_{ij}$ is a 2-form on the $(d-2)-$sphere. However, their analysis does not find any 2-form type modes on the sphere.  They classify the perturbations into three types of tensorial, vector and scalar modes depending on their behaviour on the $(d-2)-$sphere, which is parametrised, here, by $i$ type indices. The tensor modes decompose into a trace-free transverse symmetric tensor, a transverse vector and two other scalars, while the vector modes decompose into a transverse vector and a scalar.

Performing a similar decomposition of $\Phia_{ij}$ gives
\be
\Phia_{ij} = {\Phia_T}_{ij} + \eth_{[i} V_{j]},
\ee
where ${\Phia_T}_{ij}$ is transverse, i.e.
\be
\eth_{i}{\Phia_T}_{ij}=0
\ee
and $V_i$ is some vector mode.  Since transverse 2-form modes on the sphere are not found in the Ishibashi and Kodama analysis, we can only conclude that ${\Phia_T}_{ij}$ parametrise trivial metric perturbations.  An intuitive reason for why this should be the case is that to construct (symmetric) metric perturbations from ${\Phia_T}_{ij}$, we will need to contract it with a $\eth_i$ derivative, which gives zero. This leaves $V_i$, which correspond to the vector modes of Ishibashi and Kodama.

As mentioned in the introduction, this is analogous to what is found in 4d. On the Schwarzschild background, the perturbed value of the imaginary part of the Weyl scalar $\Psi_2$, which is gauge invariant and has $\Phia_{ij}$ as its higher dimensional generalisation, satisfies a decoupled equation \cite{price}. Furthermore, this decoupled equation is equivalent to the Regge-Wheeler equation \cite{regge} describing vector mode (or axial) perturbations of the Schwarzschild background \cite{price}.  The real part of the Weyl scalar $\Psi_2$, which has $\Phi \equiv \ell^a n^b \ell^c n^d C_{abcd}$ as its higher dimensional generalisation, is not gauge invariant.  From this, one can construct a gauge invariant quantity using metric perturbations and show that this also decouples and the decoupled equation is equivalent to the Zerilli equation \cite{zerilli} describing scalar mode (or polar) perturbations of the Schwarzschild background \cite{aksteiner}.  Analogously, $\Phi$ is not gauge invariant on a Schwarzschild-Tangherlini background.  However, we do not attempt to construct a gauge invariant quantity from this since we are only considering gauge invariant quantities constructed from the Weyl tensor.  Had we done so, we would presumably find that this satisfies the scalar mode master equation of Ishibashi and Kodama \cite{ishi}. 

In conclusion, we find that the only gauge invariant quantities constructed from Weyl tensor components that decouple on the Schwarzschild background are $\Phia_{ij}$. Furthermore, these components are shown to be related to the vector modes of Ishibashi and Kodama \cite{ishi}.

\section{Constructing solutions of perturbation equations} \label{hertz}

The existence of a map in 4d for constructing exact linear perturbations given the existence of a decoupled equation was demonstrated, given some technical assumptions, by Kegeles and Cohen \cite{kc, kc2} and Chrzanowski \cite{chr}.  However, Wald put this map on a firmer basis with his elegant proof of it \cite{wald}.  In this section, we review Wald's result \cite{wald} regarding the construction of solutions of perturbation equations, given the existence of decoupled equations, emphasising the generality of his argument.

We consider as our background, a solution of the vacuum Einstein equation
\be
\bar{R}_{a b}=\Lambda \bar{g}_{a b}.
\ee
We would like to find a solution of some linear (perturbation) equation
\be
\mathcal{E}(f)=0
\ee
on this background, where $f$ is the perturbation of some field.  For example, for an electromagnetic perturbation, $f$ is the 1-form potential $A$ and the equation above reduces to the Maxwell equation
\be
\na^a (\na_a A_b - \na_b A_a)=0.
\ee
In the case of gravitational perturbations, $f$ is the associated metric perturbation, usually denoted by $h$, while $\mathcal{E}$ is the linearised Einstein operator
\be \label{eqn:gravpert}
[\mathcal{G}^{(1)}(h)]_{a b}=\na_c \na_{(a} {h_{b)}}^c -\frac{1}{2}(\na^2 h_{ab}+\na_a \na_b h)-\frac{1}{2}\bar{g}_{ab}\left( \na_c \na_d h^{cd} - \na^2 h \right) + \frac{1}{2} \Lambda \left( h \bar{g}_{ab} - d h_{ab}\right),
\ee
where all indices above have been raised using the background metric $\bar{g}$, $\na$ is the Levi-Civita connection of $\bar{g}$ and $h$ is the trace of $h_{ab}$.

Now, assume that we can construct another equation, which is satisfied by a quantity $\mathcal{T}(f)$, 
\be \label{gendecoupeqn}
\mathcal{O}\left( \mathcal{T}(f) \right)=0,
\ee
where $\mathcal{O}$ and $\mathcal{T}$ are some linear operators.  Such an equation, where $\mathcal{T}(f)$ is more simple than $f$, is usually referred to as a decoupled equation.  In 4d, a decoupled equation for electromagnetic and gravitational perturbations can only be derived for vacuum solutions admitting a null geodesic, shear-free congruence \cite{teuk,teuk1}.  By the Goldberg-Sachs theorem \cite{goldberg}, this is equivalent to the condition that the solution be algebraically special (i.e. Petrov type II or more special).

In higher dimensions, the existence of a decoupled equation for electromagnetic and gravitational perturbations places more restrictive conditions on the background solution \cite{hdecoup}.  For $d>4$, the background solution must admit a null geodesic congruence with vanishing optics, i.e. the background solution must be Kundt.  Kundt solutions are necessarily of CMPP \cite{cmpp} type II \cite{highkundt}.

The existence of a decoupled equation implies the existence of a linear operator $\mc S$ such that
\be \label{opidentity}
\mc {S E}=\mc {O T}.
\ee
Simply put, operator $\mc S$ describes how the decoupled equation is derived from the perturbation equation.

The key idea in Wald's derivation is the use of operator adjoints.  Given a linear differential operator $\mc O$ acting on a tensor field $t \in T$ and taking it to a tensor field $u \in U$, the unique adjoint operator $\mc O^\dag$ is defined via
\be
(u,\mc O t) = (\mc O^\dag u, t)
\ee
up to a total divergence, where the inner product between any tensor fields $v_1$ and $v_2$ of rank $m$ is defined by
\be
(v_1, v_2) \equiv \int {v_1}^{a_1 \ldots a_{m}} {v_2}_{a_1 \ldots a_{m}} d(Vol).
\ee

We proceed by taking the adjoint of the operator equivalence \eqref{opidentity} and using the fact that $(\mc{O}_1 \mc O_2)^\dag=\mc O_2^\dag \mc O_1^\dag$ to get
\be
\mc{E^\dag S^\dag}=\mc{T^\dag O^\dag}.
\ee 
The operator equivalence above implies that given a solution $\psi$ of
\be \label{genhertzeqn}
\mc O^\dag \psi=0, 
\ee
$\mc S^\dag \psi$ solves the equation
\be
\mc E^\dag(f)=0.
\ee
For all cases of interest in this paper, the operator $\mc E$ is self-adjoint, i.e. $\mc E^\dag=\mc E$.  Thus,
\be \label{pertsol}
\mc O^\dag \psi=0 \implies \mc E(\mc S^\dag \psi)=0.
\ee
The solution $\psi$ is called the Hertz potential.  Furthermore,
\be
\mc{O (T S^\dag}\psi)=\mc{(O T)} S^\dag \psi=\mc{(S E)} S^\dag \psi=\mc S \mc E (S^\dag \psi)=0,
\ee
where we have used \eqref{opidentity} in the third equality and \eqref{pertsol} in the final equality.  Hence, $\mc{T S}^\dag \psi$ solves the decoupled equation
\be
\mc O \varphi =0.
\ee

\subsection{Electromagnetic perturbations of Kundt solutions} \label{elec}

In \cite{hdecoup}, it was shown that for electromagnetic perturbations, a higher dimensional analogue of the decoupled equation found in 4d by Teukolsky \cite{teuk1} can only be derived for Kundt solutions.  The higher dimensional decoupled equation is analogous to the Teukolsky equation in the sense that in both cases $\mc T(A)$ is the highest boost weight component of the Maxwell field $F=d A$. Hence,
\be
\mc T_E(A)_i \equiv \varphi_i = \ell^a \, {m_i}^b \, (\na_a A_b - \na_b A_a).
\ee

The decoupled equation is
\be
\mc O_E (\varphi)_i=0,
\ee
where
\be 
  \mc O_E(\varphi)_i=\left(2\tho'\tho + \eth_j\eth_j + \rho'\tho -4\tau_j\eth_j + \Phi-\tfrac{2d-3}{d-1} \, \Lambda \right)\varphi_i + 2(-2\tau_{[i}\eth_{j]} + \Phis_{ij} + 2\Phia_{ij})\varphi_j . \label{Oe}
\ee

As noted before,
\be \label{elecpert}
\mc E_E(A)_b=\na^a (\na_a A_b - \na_b A_a)=0.
\ee
Operator $\mc S_E$ is found by considering the derivation of the decoupled equation in \cite{hdecoup}.  To simplify the derivation of $\mc S_E$, we assume from the beginning that the background under consideration is Kundt, i.e. $\kappa^{(0)}={\rho^{(0)}}_{ij}=0$.

Also, in the derivation in \cite{hdecoup}, a vector potential $A$ is not introduced.  Hence, the Bianchi identity $dF=0$ is non-trivial.  Here we let $F=dA$.  Thus, of the Maxwell equations (4.4)--(4.7) in \cite{hdecoup}, only (4.4), (4.6) and their associated primed equations are non-trivial:
\begin{align}
(4.4) & \iff -\ell^a \,  \mc E_E(A)_a=0, \\
(4.4)' & \iff - n^a \, \mc E_E(A)_a=0, \\
(4.6) \iff (4.6)' &  \iff {m_i}^a \, \mc E_E(A)_a=0.
\end{align}

The derivation begins by considering the combination $\tho (4.6) + \eth_j \left[ \delta_{ij} (4.4)-(4.5) \right]$, which results in equation (4.9) of \cite{hdecoup} \footnote{We use the notation of \cite{hdecoup} to denote components of the Maxwell field strength; that is $\vphi_i=\ell^a \mb i^b F_{ab}$, $F=\ell^a n^b F_{ab}$, $F_{ij}=\mb i^a \mb j^b F_{ab}$ and $\vphi_i'=n^a \mb i^b F_{ab}$.}
\begin{eqnarray}
  0 &=& (2\tho' \tho + \eth_j \eth_j) \vphi_i + 2[\tho,\tho']\vphi_i - 
         [\tho,\eth_j] (F_{ij}+F\del_{ij}) + [\eth_i,\eth_j] \vphi_j \nn \\
    &&  + \tho \big(-(2\rho'_{[ij]}-\rho' \del_{ij}) \vphi_j
                      + 2(F_{ij}+F\del_{ij}) \tau_j \big)\label{eqn:maxmaster}\\
    &&  - \eth_i \big(\tau'_j \vphi_j 
                    \big)+ \eth_j \big(2\tau'_{[i} \vphi_{j]}\big),\nn
\end{eqnarray}
where, as noted before, we assume $\kappa=\rho_{ij}=0$. This is equivalent to considering
\be \label{eqn:maxmasterx}
\tho \left( {m_i}^a \, \mc E_E(A)_a\right)  -\eth_i \left( \ell^a \,  \mc E_E(A)_a\right).
\ee

The next manipulation that involves equations (4.4) and (4.6) is eliminating the combination $\tho(F_{ij}+F \del_{ij})$.  This is done by adding $-(2\tau_j+\tau'_j)(\delta_{ij} (4.4)-(4.5))$ to \eqref{eqn:maxmaster} or equivalently \eqref{eqn:maxmasterx}.  Hence, we have
\be \label{eqn:maxmasterx1}
\tho \left( {m_i}^a \, \mc E_E(A)_a\right)  -\eth_i \left( \ell^a \,  \mc E_E(A)_a\right) + (2\tau_i+\tau'_i)\left( \ell^a \,  \mc E_E(A)_a\right) .
\ee

The rest of the derivation does not make use of the Maxwell equation, that is it involves either the Bianchi identity or the Newman-Penrose equations.  Hence, operator $\mc S_E$ is given by equation \eqref{eqn:maxmasterx1}
\be \label{Se}
\mc S_E(J)_i=\tho({m_i}^a J_a)-(\eth_i-2\tau_i-\tau'_i)(\ell^a J_a).
\ee

The Hertz potential $\psi_H$ is given by solving
\be \label{hertzeq}
\mc {O}_E^\dag ({\psi_H})_i =0.
\ee
We derive $\mc O_E^\dag$ by considering the inner product
\be
\left(\psi_i, \mc O_E (\vphi)_i \right).
\ee
For the inner product to be well-defined $\psi_i$ must be boost weight -1 since $\mc O_E (\vphi)_i$ is boost weight 1.
Using equation \eqref{Oe}
\begin{align}
(\psi_i, \mc O_E (\vphi)_i)=\left( \left[2 \tho'^\dag \tho^\dag +{\eth_j}^\dag {\eth_j}^\dag + \tho^\dag \rho' + -4 {\eth_j}^\dag \tau_j \Phi -\textstyle{\frac{2d-3}{d-1}} \Lambda \right] \psi_i \right. \notag \\
 + \left. 2 \left[ -2 {\eth_{[i}}^\dag \tau_{j]} + \Phis_{ij} -2 \Phia_{ij} \right] \psi_j , \vphi_i \right) 
\end{align}
Using equations \eqref{thoad}--\eqref{ethad} for the adjoint of the GHP covariant derivatives, commutator \eqref{C1} and the NP equation \eqref{NP4} gives
\begin{align} \label{Oedag}
 \mc {O}_E^\dag ({\psi})_i=\left(2\tho'\tho + \eth_j\eth_j + \rho'\tho + \Phi-\tfrac{ \Lambda}{d-1} \right)\psi_i + 2 (-2\tau_{[i}\eth_{j]} + \Phis_{ij} - 2\Phia_{ij})\psi_j . 
\end{align}

To find a solution of the electromagnetic perturbation equation \eqref{elecpert}, we also need to calculate the adjoint of $\mc S$.  We do this by considering the inner product 
\be
(\psi_i, \mc S(A)_i)=\left( \left[ {m_i}^a \tho^\dag + \ell^a (-{\eth_i}^\dag +2\tau_i +\tau') \right] \psi_i , A_a\right)
\ee
where we have used \eqref{Se}.  Using equations \eqref{thoad} and \eqref{ethad} gives
\be
\mc S^\dag (\psi)_a=\left[ -{m_i}_a \, \tho +\ell_a (\eth_i + \tau_i) \right] \psi_i.
\ee
Thus, using the results of section \ref{hertz}, we have that if ${\psi_H}_i$ is a solution to \eqref{hertzeq},
where $\mc O_E^\dag$ is given in equation \eqref{Oedag}, then
\be
\mc S_E^\dag (\psi_H)_a=\left[ -{m_i}_a \, \tho +\ell_a (\eth_i + \tau_i) \right] {\psi_H}_i
\ee
is a solution of the electromagnetic perturbation equation \eqref{elecpert}.

Also,
\be
\mc{T}_E \mc S_E^\dag (\psi_H)_i=\tho^2 (\psi_H)_i
\ee
is a solution of the decoupled equation $\mc O_E (\vphi)_i=0$.

Consider a doubly Kundt solution. This is a solution with two null geodesic congruences with vanishing optics, i.e. we also have $\kappa'^{(0)}_i=\rho'^{(0)}_{ij}=0$. In this case, the boost weight -1 component of the Maxwell field $\vphi'$ also satisfies a decoupled equation---the prime of the decoupled equation \eqref{Oe}
\be
\mc O'_E(\vphi')_i=0,
\ee
where (using equation \eqref{Oe} and commutator \eqref{C1})
\begin{align}
{\mc O'}_E(\vphi')_i=\left(2\tho'\tho + \eth_j\eth_j -2(\tau_j+\tau'_j)\eth_j +2\tau_j \tau'_j - \Phi-\tfrac{2d-3}{d-1} \, \Lambda \right)\varphi'_i \notag \\
+ 2 \left(-2\tau'_{[i}\eth_{j]} + \Phis_{ij} + 4\Phia_{ij}+2\tau'_{[i}\tau_{j]}\right) \varphi'_j .
\end{align}

In this case, if ${\psi'_H}_i$ is a solution to the prime of equation \eqref{hertzeq}, then $\mc {S'_E}^\dag (\psi'_H)_a$ is also a solution of the perturbation equation.

\subsection{Gravitational perturbations of Kundt solutions} \label{grav}

For gravitational perturbations we assume that the perturbed solution also satisfies the vacuum Einstein equation. The metric perturbation $h$ satisfies
\be \label{lineineq}
\mc E_G (h)_{ab}=0,
\ee 
where $\mc E_G$ is given in equation \eqref{eqn:gravpert}.

In \cite{hdecoup}, it was shown that as with electromagnetic perturbations an analogue of the Teukolsky decoupled equation for gravitational perturbations \cite{teuk1} only exists for Kundt solutions.  The decoupled equation is solved by the boost weight +2 components of the perturbed Weyl tensor $\Omega_{ij}$, which generalises the complex Weyl scalar $\Psi_0$ of the 4d NP formalism to higher dimensions.
\begin{align}
\mc T_G(h)_{ij} &\equiv \Omega_{ij} = -\frac{1}{2} \left\lbrace \left(\eth_{(i} \eth_{j)} - \displaystyle{\frac{\delta_{ij}}{d-2}} \eth_{k} \eth_{k} \right)  -2 \left(\tau'_{(i} \eth_{j)} - \displaystyle{\frac{\delta_{ij}}{d-2}} \tau'_{k} \eth_{k} \right) - \left(\rho'_{ij} - \displaystyle{\frac{\delta_{ij}}{d-2}} \rho'\right) \tho \right. \notag \\
& \left. -2 \left(\tho \rho'_{ij} - \displaystyle{\frac{\delta_{ij}}{d-2}} \tho \rho' \right) \right\rbrace (\ell^a \ell^b h_{ab}) - \displaystyle{\frac{1}{2}} \tho^2 \left\lbrace  \left({m_{i}}^a {m_{j}}^b- \displaystyle{\frac{\delta_{ij}}{d-2}} {m_{k}}^a {m_{k}}^b\right) h_{ab} \right\rbrace  \notag \\
& + \left\lbrace \tho \eth_{(i} - \tau'_{(i} \tho - (\tho \tau'_{(i}) \right\rbrace (\ell^a {m_{j)}}^b h_{ab})
- \displaystyle{\frac{\delta_{ij}}{d-2}} \left\lbrace \tho \eth_{k} - \tau'_{k} \tho - (\tho \tau'_{k}) \right\rbrace (\ell^a {m_{k}}^b h_{ab}).
\end{align}
The decoupled equation is
\be \label{decoupgrav}
\mc O_G(\Omega)_{ij}=0,
\ee
where
\begin{align} \label{eqn:gravperts}
 \mc O_G(\Omega)_{ij} = \left(2\tho'\tho+ \eth_k \eth_k + \rho'\tho - 6\tau_k\eth_k  + 4\Phi - \tfrac{2d}{d-1} \La \right) \Om_{ij} & \notag \\
 + 4  \left(\tau_k\eth_{(i|}- \tau_{(i|}\eth_k + \Phis_{(i|k} + 4\Phia_{(i|k}\right) & \Om_{k|j)} + 2\Phi_{ikjl} \, \Om_{kl}.
\end{align}

To derive operator $\mc S_G$, we assume that the gravitational perturbation also generates a first order energy-momentum tensor.  Then, $\mc S_G$ is the operator acting on the first order energy-momentum tensor in the inhomogeneous decoupled equation
\be
\mc O_G (\Om)_{ij}=8\pi \, \mc S_G \left( T^{(1)}_{ab}\right) _{ij}.
\ee
Therefore, we need to rederive the decoupling result of Ref. \cite{hdecoup} assuming a non-zero first order energy-momentum tensor.  Thus, we must use the more general ({\it{viz.}} including matter) NP equations, Bianchi identities and commutator relations found in Ref. \cite{ghp}.  In order, to simplify the calculation, we may assume from the onset that the background is Kundt.

Going through the derivation of the decoupled equation for gravitational perturbations given in Ref. \cite{hdecoup}, except using the more general equations that include matter terms gives
\begin{align}
 \mc S_G \left(T\right)_{ij}=\tfrac{1}{d-2} \, \del_{ij} \left( 2\tho'\tho+ \eth_k \eth_k + \rho'\tho - 6\tau_k\eth_k  + 4\Phi \right) (\ell^a \ell^b T_{ab}) - \left( \Phis_{ij} - \tho \rho'_{(ij)} \right)  (\ell^a \ell^b T_{ab}) \notag \\
+2\left( \tho \eth_{(i} - (2\tau_{(i}+\tau'_{(i}) \tho -(\tho \tau'_{(i}) \right)(\ell^a {m_{j)}}^b T_{ab}) -\tho \tho ({m_{i}}^a {m_{j}}^b T_{ab}) + \tfrac{1}{d-2} \, \del_{ij} \, \tho \tho (g^{ab} T_{ab}).
\end{align}

The adjoints of $\mc O_G$ and $\mc S_G$ can be derived in a fashion analogous to the electromagnetic case
\begin{align} \label{Odagg}
 \mc O_G^\dag (\Pi)_{ij} = \left(2\tho'\tho+ \eth_k \eth_k + \rho'\tho +2\tau_k\eth_k  + 4\Phi + \tfrac{2(d-4)}{d-1} \La \right) \Pi_{ij} & \notag \\
 + 4  \left(\tau_k\eth_{(i|}- \tau_{(i|}\eth_k + \Phis_{(i|k} - 4\Phia_{(i|k}\right) & \Pi_{k|j)} + 2\Phi_{ikjl} \, \Pi_{kl},
\end{align}

\vspace{-5mm}

\begin{align} \label{SdagG}
 \mc S_G^\dag \left(\Pi\right)_{ab}=- \ell_a \ell_b \left( \Phis_{ij} + \rho'_{(ij)} \tho \right)  \Pi_{ij} 
+2 \ell_{(a}m_{|j| b)} \left( \tho \eth_{i} + (\tau_{i}+\tau'_{i}) \tho \right)\Pi_{ij} - {m_{i}}_a {m_{j}}_b \, \tho^2 \, \Pi_{ij}.
\end{align}

Thus, given a Hertz potential $\Omega_H$ that is a solution of
\be \label{Pieqn}
\mc O_G^\dag (\Omega_H)_{ij}=0,
\ee 
where operator $\mc O_G^\dag$ is given in eq. \eqref{Odagg}, then
\be
\mc S_G^\dag \left(\Omega_H \right)_{ab}
\ee
where $\mc S_G^\dag$ is given in eq. \eqref{SdagG}, is a solution of the gravitational perturbation equation \eqref{lineineq}.

Furthermore,
\be
\mc T_G \mc S_G^\dag \left(\Omega_H \right)_{ij} = \displaystyle{\frac{1}{2}} \tho^4 {\Omega_H}_{ij}
\ee
is a solution of the decoupled equation \eqref{decoupgrav}.

In the case of a background that is doubly Kundt, the prime of all the above equations hold also.

For the case of doubly Kundt solutions, the operators $\mc O$, $\mc O'$, $\mc O^\dag$ and $\mc O'^\dag$ encode the same physical information.  Therefore, there should be some connection between them.  Indeed, in 4d, for type D solutions, it is known that the solution of the equations corresponding to the above operators are related by various factors of the background value of the Weyl scalar $\Psi_2$ (see Refs. \cite{teuk3, chand, wald} and references therein).  It is very simple to prove such relations using the Bianchi identities.  However, in higher dimensions the situation is more complicated.  We have not been able to derive similar expressions relating the solutions to the various equations defined by the four operators above.  

\subsection{Asymptotic behaviour of metric perturbations of near horizon geometry of 5d cohomogeneity-1 extreme Myers-Perry solutions}

In this section, we consider as an application of the Hertz potential map developed above, the asymptotic behaviour of the metric perturbation of the near horizon geometry of 5d cohomogeneity-1 extreme Myers-Perry black hole.  This consideration is motivated by a recent proposal that quantum gravity on the near horizon of a class of 5d solutions of which the above solution is an example, with appropriate asymptotic fall-off conditions on the metric perturbation is equivalent to a CFT, which can be used to calculate the Bekenstein-Hawking entropy of the original solution \cite{guica5d}. Thus, giving a statistical counting of the black hole's degrees of freedom.

Turning off the graviphoton charge in the solution discussed in \cite{guica5d} gives the 5d cohomogeneity-1 extreme Myers-Perry black hole solution with near-horizon geometry
\be \label{NHMP}
ds^2=\frac{r_{+}^2}{4} \left\lbrace  -R^2 dT^2 +\frac{dR^2}{R^2} +2 (d\psi + \cos \theta d\phi + R dT)^2 + d\Omega_{(2)}^2 \right\rbrace ,
\ee
where $r_{+}$ is the event horizon radius and $d\Omega_{(2)}^2$ is the round metric on $S^2$.  This (doubly Kundt) solution is studied in Ref. \cite{durkee10}, where the decoupling result of \cite{hdecoup} is used to predict an instability of the corresponding extreme Myers-Perry solution.
Choose the following null frame
\begin{gather}
 \ell=\frac{r_{+}}{2\sqrt{2}}\left( -R \, dT+\frac{dR}{R} \right), \quad n=\frac{r_{+}}{2\sqrt{2}}\left(R \, dT+\frac{dR}{R} \right), \notag\\
 m_2=\frac{r_{+}}{\sqrt{2}} (d\psi + \cos \theta d\phi + R dT), \quad m_{\alpha}=\frac{r_{+}}{2} \hat{e}_{\alpha},
\end{gather}
where $\alpha=3, \, 4$ and $\hat{e}_{\alpha}$ form an orthonormal basis on $S^2$.

In Ref. \cite{durkee10}, it was shown that for the geometry with metric \eqref{NHMP} with basis chosen as above
\begin{gather}
 \kappa_i=\kappa'_i=0, \quad \rho_{ij}=\rho'_{ij}=0, \quad \tau_i+\tau'_i=0, \notag \\ \label{backcur}
\Omega_{ij}= \Omega'_{ij}=0, \quad \Psi_{ijk}=\Psi'_{ijk}=0, \quad \Phi_{ijkl}=\hat{R}_{ijkl}, \quad \Phia_{ij}=-\frac{4}{r_{+}^2} (m_3 \wedge m_4)_{ij},
\end{gather}
where $\hat{R}_{ijkl}$ is the Riemann tensor of the three dimensional space $\mc H$ with metric
\be
ds_{\mc H}^2=\frac{r_{+}^2}{4} \left\lbrace 2 (d\psi + \cos \theta d\phi)^2 + d\Omega_{(2)}^2 \right\rbrace.
\ee

In order to determine the asymptotic behaviour of the metric perturbation of the near horizon geometry with metric \eqref{NHMP}, we must first solve the Hertz potential equation
\be
\mc O_G^\dag (\Pi)_{ij}=0,
\ee
where operator $\mc O_G^\dag$ is given in \eqref{Odagg}.  Then,
\be \label{hab}
h_{ab}= \frac{1}{2} \, \ell_a \ell_b ( \hat{R}_{ij}  \Pi_{ij}) + 2 \ell_{(a}m_{|j| b)} \tho \eth_{i} \Pi_{ij} - {m_{i}}_a {m_{j}}_b \, \tho^2 \, \Pi_{ij}
\ee
is a solution of the gravitational perturbation equation in the ingoing radiation gauge, where we have used eq. \eqref{SdagG} and eqs. \eqref{backcur} above.  Space $\mc H$ is not an Einstein solution.  Thus, the $\ell_a \ell_b$ component of $h_{ab}$ is non-zero.

Assume the following separability ansatz for $\Pi_{ij}$
\be
\Pi_{ij}=\chi(T,R) Y_{ij}(\theta, \phi, \psi).
\ee
Substituting this ansatz into the equation for $\Pi_{ij}$, i.e. eq. \eqref{Pieqn} gives \footnote{The steps involved in this computation are almost identical to that given in appendix A of Ref. \cite{durkee10}.}
\be
(D^2-q^2-\lambda)\chi=0, \qquad \mc O^{(2)} Y_{ij}=\lambda Y_{ij},
\ee
where $D$ is some charge covariant derivative for some $AdS_2$ scalar with charge $q$ defined in Ref. \cite{durkee10} and $\mc O^{(2)}$ is some operator given in Ref. \cite{durkee10}, which we do not need to know about in detail here.  Hence, $\chi$ solves the equation for a massive, charged scalar in an $AdS_2$ background with homogeneous electric field.  Such an equation has been studied by a number of authors \cite{strom99, amsel, diaskerr, durkee10}.  At large $R$,
\be
\chi \sim R^{-\Delta_{\pm}}, \quad \Delta_{\pm}=\frac{1}{2}\pm \sqrt{\lambda + \frac{1}{4}}.
\ee

Hence, from the form of $h_{ab}$ given in \eqref{hab} we can conclude that for large $R$
\be
h_{ab} \sim R^{\textstyle{\frac{1}{2}}\pm \textstyle{\frac{1}{2}} \eta}
\begin{pmatrix}
\vspace{1.5mm}
 \mc O(1) & \underline{\mc O(\textstyle{\frac{1}{R^2}})} & \mc O(\textstyle{\frac{1}{R}}) & \underline{\mc O(\textstyle{\frac{1}{R}})} \\ \vspace{1.5mm}
& \underline{\mc O(\textstyle{\frac{1}{R^3}})} & \mc O(\textstyle{\frac{1}{R^3}}) & \mc O(\textstyle{\frac{1}{R^3}}) \\
\vspace{1.5mm}
& & \mc O(\textstyle{\frac{1}{R^2}}) & \underline{\mc O(\textstyle{\frac{1}{R}})} \\
& & & \underline{\mc O(\textstyle{\frac{1}{R}})}
\end{pmatrix}, \qquad \eta=\sqrt{1+4\lambda}.
\ee
where the columns and rows indicate the $T$, $R$, $\psi$ and $\theta$ and $\phi$ (collectively labelled $\alpha$) components, respectively.  The same result would be found if we considered $n$ as the WAND of choice, that is if $h_{ab}$ were in outgoing gauge.

Comparing this with the fall-off conditions in Ref. \cite{guica5d}, we find that the $TR$, $T\psi$, $RR$, $\psi \alpha$ and $\alpha \beta$ components are the most restrictive.  These components have been underlined in the matrix above.  Thus, in order to satisfy the fall-off conditions, $\eta$ must be real. We must also choose the lower sign (corresponding to normalisable modes) and require that $\eta \geq 1$ or $\lambda \geq 0$.  Recall that $\lambda$ is the eigenvalue of operator $\mc O^{(2)}$.  The spectrum of operator $\mc O^{(2)}$ was studied in appendix B of Ref. \cite{durkee10}.  It is clear from the study of gravitational scalar modes that there exists modes for which
\be
\lambda=2+\kappa(\kappa+1)+|m|(\kappa+\textstyle{\frac{1}{2}})-m^2/8,
\ee
where $m$ is an integer and $\kappa$ is a positive integer.  For $\kappa=1$, $m \geq 15$, $\lambda < 0$.  Hence, there exist modes that violate the fall-off conditions.  It was shown in Ref. \cite{durkee10} that for all axisymmetric modes ($m=0$) $\lambda \geq 0$.  Hence, all axisymmetric modes satisfy the boundary conditions.

There is a similar proposal for the entropy counting of the 4d extremal Kerr solution \cite{kerrcft}.  In Ref. \cite{kerrcft}, as in the 5d case, the metric perturbation of the NHG is assumed to satisfy a certain asymptotic form.  The asymptotic behaviour of the NHG of the extremal Kerr solution has been studied in \cite{amsel} and \cite{diaskerr}.  The results we find for the 5d case are the same as they found in the 4d case. 

\paragraph*{Acknowledgements}
I would like to thank Harvey Reall for suggesting this project and reading through a draft manuscript.  His many comments and suggestions have been invaluable throughout. I thank Oscar Dias for helpful discussions. I am supported by an EPSRC grant. Also, I thank St. John's College Cambridge for their support.

\newpage
\appendix

\section{Higher dimensional GHP formalism} \label{app:ghp}

In this appendix, we review the higher dimensional GHP formalism of \cite{ghp}.  Given a background solution, we choose a null frame $(\ell, n ,\mb i)$ such that in this frame, the metric takes the form
\be \label{metric}
g_{ab}=2 \ell_{(a} n_{b)}+ \mb i_{a} \mb i_{b}.
\ee

In the GHP formalism, one breaks complete covariance by singling out two null directions ($\ell$ and $n$) at each point, but preserves covariance in the remaining directions.  This is in contrast to the NP formalism where none of the covariance is preserved.

At any point, the Lorentz group divides into
\begin{itemize}
 \item boosts ($\lambda$ a real function):
\be
\ell \rightarrow \lambda \, \ell, \quad n \rightarrow \lambda^{-1} n, \quad \mb i \rightarrow \mb i, \label{app:boost}
\ee
 \item spins ($X_{ij} \in SO(d-2)$): 
\be
\ell \rightarrow \ell, \quad n \rightarrow n, \quad \mb i \rightarrow X_{i j} \mb j, \label{app:spin}
\ee
\item null rotations about $\ell$ ($z_i$ $d-2$ real functions):
\be
\ell \rightarrow \ell, \quad n \rightarrow n + z_i \mb i - \half z^2 \ell, \quad \mb i \rightarrow \mb i - z_i \ell, \label{app:rotl}
\ee
\item null rotations about $n$ ($z_i$ $d-2$ real functions):
\be
\ell \rightarrow \ell + z_i \mb i - \half z^2 n, \quad n \rightarrow n, \quad \mb i \rightarrow \mb i - z_i n, \label{app:rotn}
\ee 
\end{itemize}
where $\lambda \neq 0$ and $X_{ij}$ is some position-dependent orthogonal matrix.

We would like to keep the subgroup that preserves the null directions, i.e. the subgroup given by boosts and spatial rotations, or spins.  Thus, we would like to work with objects that transform covariantly under this subgroup.

Define any scalar $\eta_{i_1 \ldots i_{s}}$ that transforms covariantly as
\be
 \eta_{i_1 \ldots i_{s}} \rightarrow \lambda^{b} \eta_{i_1 \ldots i_{s}}
\ee
under boosts and 
\be
 \eta_{i_1 \ldots i_{s}} \rightarrow X_{i_1 j_1} \cdots X_{i_s j_s} \eta_{j_1 \ldots j_{s}}
\ee
under spins, a GHP scalar of boost weight $b$ and spin $s$.  Evidently, the product of two GHP scalars of boost weights $b_1$ and $b_2$ and spins $s_1$ and $s_2$, respectively, gives a GHP scalar of boost weight $b_1+b_2$ and spin $s_1+s_2$.

Defining the covariant derivatives of the basis vectors as
\be
L_{a b} = \nabla_{b} \ell_a, \quad N_{a b} = \nabla_{b} n_a, \quad \M i_{a b} = \nabla_{b} \mb i_a,
\ee
one finds that not all the scalars formed from the projection of these objects into the basis are GHP scalars.  Those that are GHP scalars are listed in table \ref{tab:npsca} \cite{ghp}.

\begin{table}[ht]
\caption{GHP scalars constructed from covariant derivatives of the basis vectors.} \centering
\vspace{1.5mm}
\begin{tabular}{c c c c l}
\hline
Spin coefficient & GHP notation & Boost weight $b$ & Spin $s$ & Interpretation\\ [1mm]
\hline 
$L_{ij}$  & $\rho_{ij}$  & 1  & 2 & expansion, shear and twist of $\lb$\\[1mm]
     $L_{ii}$  & $\rho=\rho_{ii}$  & 1  & 0 & expansion of $\lb$\\[1mm]
    $L_{i0}$  & $\kap_{i}$   & 2  & 1 & non-geodesity of $\lb$\\[1mm]
    $L_{i1}$  & $\tau_{i}$   & 0  & 1 & transport of $\lb$ along $n$\\[1mm]
    $N_{ij}$  & $\rho'_{ij}$ & -1 & 2 & expansion, shear and twist of $n$\\[1mm]
     $N_{ii}$  & $\rho'=\rho'_{ii}$ & -1 & 0 & expansion of $n$\\[1mm]
     $N_{i1}$  & $\kap'_{i}$  & -2 & 1 & non-geodesity of $n$\\[1mm]
    $N_{i0}$  & $\tau'_{i}$  & 0  & 1 & transport of $n$ along $l$\\[1mm]\hline
\end{tabular}
\label{tab:npsca}
\end{table}

Notice that we have used a prime operation, which interchanges the null basis vectors
\be
' \ : \  \ell \leftrightarrow n.
\ee
The prime operation is especially useful when considering type D backgrounds, since in this case $\ell$ and $n$ are essentially equivalent.

The non-GHP covariant scalars formed from the covariant derivative of the basis vectors can be used to construct GHP covariant derivatives .  For a GHP scalar $\eta_{i_1 \ldots i_{s}}$ of boost weight $b$ and spin $s$, we define its GHP covariant derivatives to be \footnote{Symbols $\tho$ and $\eth$, pronounced ``thorn'' and ``eth'', respectively are old Germanic letters that have been retained in the Icelandic alphabet.}
 \begin{eqnarray} \label{ghpdertho}
    \tho T_{i_1 i_2...i_s} &\equiv & \ell \cdot \partial T_{i_1 i_2...i_s} - b L_{10} T_{i_1 i_2...i_s} 
                                     + \sum_{r=1}^s \M{k}_{i_r 0} T_{i_1...i_{r-1} k i_{r+1}...i_s},\\
    \tho' T_{i_1 i_2...i_s} &\equiv & n \cdot \partial T_{i_1 i_2...i_s} - b L_{11} T_{i_1 i_2...i_s} 
                                     + \sum_{r=1}^s \M{k}_{i_r 1} T_{i_1...i_{r-1} k i_{r+1}...i_s}, \label{ghpdertho'} \\ \label{ghpderm}
    \eth_i T_{j_1 j_2...j_s} &\equiv & \mb i \cdot \partial T_{j_1 j_2...j_s} - b L_{1i} T_{j_1 j_2...j_s} 
                                     + \sum_{r=1}^s \M{k}_{j_r i} T_{j_1...j_{r-1} k j_{r+1}...j_s}.
  \end{eqnarray}

In GHP notation, the Newman-Penrose, Bianchi and the commutator equations are much more compact that in the NP formalism \cite{highNP}. For convenience, we write these equations, here, for an Einstein spacetime \cite{hdecoup}.

\subsection{Newman-Penrose equations}
\newcounter{oldeq}
\setcounter{oldeq}{\value{equation}}
\renewcommand{\theequation}{NP\arabic{equation}}
\setcounter{equation}{0}
\begin{eqnarray}\label{NP1}
  \tho \rho_{ij} - \eth_j \kap_i &=& - \rho_{ik} \rho_{kj} -\kap_i \tau'_j - \tau_i \kap_j - \Om_{ij},\\[3mm]
  \tho \tau_i - \tho' \kap_i &=& \rho_{ij}(-\tau_j + \tau'_j) - \Psi_i,\label{NP2}\\[3mm]
  2\eth_{[j|} \rho_{i|k]}     &=& 2\tau_i \rho_{[jk]} + 2\kap_i \rho'_{[jk]} - \Psi_{ijk} ,\label{R:ethrho}\label{NP3}\\[3mm]
  \tho' \rho_{ij} - \eth_j \tau_i &=& - \tau_i \tau_j - \kap_i \kap'_j - \rho_{ik}\rho'_{kj}-\Phi_{ij}
                                      - \tfrac{\La}{d-1}\del_{ij}.\label{NP4}
\end{eqnarray}
Another four equations can be obtained by taking the prime $'$ of these four (i.e.\ by exchanging the vectors $\lb$ and $\nb$).
\renewcommand{\theequation}{A.\arabic{equation}}
\setcounter{equation}{\value{oldeq}}

\subsection{Bianchi equations}\label{sec:bianchi}
\setcounter{oldeq}{\value{equation}}
\renewcommand{\theequation}{B\arabic{equation}}
\setcounter{equation}{0}
{\noindent\bf Boost weight +2:}
\begin{eqnarray}
  \tho \Ps_{ijk} - 2 \eth_{[j}\Om_{k]i} 
                  &=& (2\Phi_{i[j|} \del_{k]l} - 2\del_{il} \Phia_{jk}-\Phi_{iljk})\kap_l \nn\\
                  && -2 (\Ps_{[j|} \del_{il} + \Ps_i\del_{[j|l} + \Ps_{i[j|l} 
                     + \Ps_{[j|il}) \rho_{l|k]} + 2 \Om_{i[j} \tau'_{k]},\label{B1}
\end{eqnarray}
{\bf Boost weight +1:}
\begin{eqnarray}
  - \tho \Phi_{ij} - \eth_{j}\Ps_i + \tho' \Om_{ij} 
                 &=& - (\Ps'_j \del_{ik} - \Ps'_{jik}) \kap_k + (\Phi_{ik} + 2\Phia_{ik} + \Phi \del_{ik}) \rho_{kj} 
                      \nonumber\\
                 &&  + (\Ps_{ijk}-\Ps_i\del_{jk}) \tau'_k - 2(\Ps_{(i}\del_{j)k} + \Ps_{(ij)k}) \tau_k 
                     - \Om_{ik} \rho'_{kj}, \label{B2}\\[3mm]
  -\tho \Phi_{ijkl} + 2 \eth_{[k}\Ps_{l]ij}
                 &=& - 2 \Ps'_{[i|kl} \kap_{|j]} - 2 \Ps'_{[k|ij}\kap_{|l]}\nn\\
                 &&  + 4\Phia_{ij} \rho_{[kl]} -2\Phi_{[k|i}\rho_{j|l]} 
                     + 2\Phi_{[k|j}\rho_{i|l]} + 2 \Phi_{ij[k|m}\rho_{m|l]}\nn\\
                 &&  -2\Ps_{[i|kl}\tau'_{|j]} - 2\Ps_{[k|ij} \tau'_{|l]}
                     - 2\Om_{i[k|} \rho'_{j|l]} + 2\Om_{j[k} \rho'_{i|l]},
                     \label{B3}\\[3mm]
  -\eth_{[j|} \Ps_{i|kl]}
                 &=& 2\Phia_{[jk|} \rho_{i|l]} - 2\Phi_{i[j} \rho_{kl]} 
                     + \Phi_{im[jk|} \rho_{m|l]} - 2\Om_{i[j} \rho'_{kl]},\label{B4}
\end{eqnarray}
{\bf Boost weight 0:}
\begin{eqnarray}
  \tho' \Ps_{ijk} -2 \eth_{[j|}\Phi_{i|k]} 
                 &=& 2(\Ps'_{[j|} \del_{il} - \Ps'_{[j|il}) \rho_{l|k]}
                     + (2 \Phi_{i[j}\del_{k]l} - 2\del_{il}\Phia_{jk} - \Phi_{iljk}) \tau_l \nn\\
                 &&  + 2 (\Ps_i \del_{[j|l} -  \Ps_{i[j|l})\rho'_{l|k]} + 2\Om_{i[j}\kap'_{k]},
                     \label{B5}\\[3mm]
  -2\eth_{[i} \Phia_{jk]} 
                 &=& 2\Ps'_{[i} \rho_{jk]} + \Ps'_{l[ij|} \rho_{l|k]} 
                     - 2\Ps_{[i} \rho'_{jk]} - \Ps_{l[ij|} \rho'_{l|k]},\label{B6}\\[3mm]
  -\eth_{[k|} \Phi_{ij|lm]} 
                 &=& - \Ps'_{i[kl|} \rho_{j|m]} + \Ps'_{j[kl|} \rho_{i|m]} 
                     - 2\Ps'_{[k|ij} \rho_{|lm]}\nn\\
                  && - \Ps_{i[kl|} \rho'_{j|m]} + \Ps_{j[kl|} \rho'_{i|m]} 
                     - 2\Ps_{[k|ij} \rho'_{|lm]}.\label{B7}
\end{eqnarray}
Another five equations are obtained by applying the prime operator to equations (\ref{B1})-(\ref{B5}) above.
\renewcommand{\theequation}{A.\arabic{equation}}
\setcounter{equation}{\value{oldeq}}

\subsection{Commutators of derivatives}
\setcounter{oldeq}{\value{equation}}
\renewcommand{\theequation}{C\arabic{equation}}
\setcounter{equation}{0}
Acting on a GHP scalar of boost weight $b$ and spin $s$, commutators of GHP derivatives can be simplified by:
\begin{eqnarray}
[\tho, \tho']T_{i_1...i_s} 
         &=& (-\tau_j + \tau'_j) \eth_jT_{i_1...i_s} + 
                    b\left( -\tau_j\tau'_j + \kap_j\kap'_j + \Phi - \textstyle{\frac{\Lambda}{d-1}} \right)T_{i_1...i_s} \nn\\
         &&  + \sum_{r=1}^s \left(\kap_{i_r} \kap'_{j} - \kap'_{i_r} \kap_{j} 
                                  + \tau'_{i_r} \tau_{j} - \tau_{i_r} \tau'_{j} + 2\Phia_{i_r j}
                            \right) T_{i_1...j...i_s}, \label{comm:thotho}\label{C1}\\[3mm]
[\tho, \eth_i]T_{k_1...k_s}
         &=& -(\kap_i \tho' + \tau'_i\tho +\rho_{ji}\eth_j)T_{k_1...k_s}
             + b\left(-\tau'_j\rho_{ji} + \kap_j\rho'_{ji} + \Psi_i \right)T_{k_1...k_s} \nn\\
         &+&  \sum_{r=1}^s \left( \kap_{k_r}\rho'_{li} - \rho_{k_r i}\tau'_l
            + \tau'_{k_r} \rho_{li} - \rho'_{k_r i} \kap_l - \Psi_{ilk_r}\right) T_{k_1...l...k_s},
            \label{comm:thoeth}\label{C2}\\[3mm]
[\eth_i,\eth_j]T_{k_1...k_s}
         &=& \left(2\rho_{[ij]} \tho' + 2\rho'_{[ij]} \tho \right) T_{k_1...k_s}
                   + b \left(2\rho_{l[i|} \rho'_{l|j]} + 2\Phia_{ij}\right) T_{k_1...k_s}\nn\\
         && + \sum_{r=1}^s \left(2\rho_{k_r [i|} \rho'_{l|j]} + 2\rho'_{k_r [i|} \rho_{l|j]} 
                                + \Phi_{ijk_r l} + \tfrac{2\La}{d-1} \del_{[i|k_r}\del_{|j]l} 
                           \right) T_{k_1...l...k_s}. \label{comm:etheth} \label{C3}
\end{eqnarray}
The result for $[\tho', \eth_i]$ can be obtained from \eqref{comm:thoeth}$'$. 
\renewcommand{\theequation}{B.\arabic{equation}}
\setcounter{equation}{\value{oldeq}}

\section{Schwarzschild-Tangherlini solution} \label{app:schw}
The Schwarzschild-Tangherlini black hole is an example of a higher dimensional type D solution.  The Schwarzschild-Tangherlini metric in Schwarzschild coordinates is
\be
ds^2= - f(r) dt^2 +dr^2/f(r) + r^2 d\Omega_{(d-2)}^2, \qquad f(r)=1-\frac{\mu}{r^{d-3}},
\ee
where $d\Omega_{(d-2)}^2$ is the round metric on a unit radius $(d-2)-$sphere.  The WANDs of the solution are \cite{schwclass}
\be
\ell=\frac{\partial}{\partial r}, \quad n = \frac{\partial}{\partial v} + {\textstyle{\frac{1}{2}}}
 f \frac{\partial}{\partial r}.
\ee
Defining $(d-2)$ orthonormal spacelike vectors
\be
m_{i} = r \hat{e}_{i} \qquad (i=2,\cdots, d),
\ee
completes a null frame $(\ell, n, m_{i})$ for the solution.

Cartan's first equation of structure $de^\mu + {\omega^\mu}_{\nu} e^{\nu}=0$ can be used to find the optical scalars associated with WANDs $\ell$ and $n$
\begin{gather}
L_{11}=-N_{01}=- {\textstyle{\frac{1}{2}}} \partial_{r}f, \notag \\
\kappa_i=\tau_i=0, \quad \rho_{ij}=\frac{\rho}{d-2} \delta_{ij}, \quad \rho=(d-2)/r,  \notag \\
\kappa'_i=\tau'_i=0, \quad \rho'_{ij}=\frac{\rho'}{d-2} \delta_{ij}, \quad \rho'=(d-2)f/(2r).
\end{gather}
Thus, the solution is an example of a Robinson-Trautman solution, that is there exists a null geodesic congruence with vanishing shear and rotation, but non-vanishing expansion. 

The curvature tensors can be derived from Cartan's second equation of structure $R_{\mu \nu}= d\omega_{\mu \nu} + {\omega_{\mu}}^{\rho} \wedge \omega_{\rho \nu}$.  Or, alternatively, one can read off the curvature tensors from appendix A of \cite{axi}
\begin{gather}
 \Omega_{ij}= \Psi_{ijk}=0, \quad \Omega'_{ij}= \Psi'_{ijk}=0, \notag \\
\Phi_{ijkl}=\frac{-4\Phi \delta_{i[k} \delta_{l]j} }{(d-2)(d-3)}  , \quad \Phis_{ij}=-{\textstyle \frac{1}{2}} \Phi_{i k j k}, \quad \Phia_{ij}=0, \quad \Phi=-\frac{(d-2)(d-3)\mu}{2r^{d-1}} .
\end{gather}
In particular,
\be
\Phis_{ij}-\frac{\Phi }{d-2}\delta_{ij}=0.
\ee
\renewcommand{\theequation}{C.\arabic{equation}}
\setcounter{equation}{\value{oldeq}}

\section{Derivation of equation \eqref{rphitdecoup}} \label{app:deriv}
This appendix is dedicated to the derivation of equation \eqref{rphitdecoup}.

We proceed by deriving an equation in which second order derivatives act only on $\Phis_{ij}- { \textstyle \frac{\Phi }{d-2}} \delta_{ij}$.  To simplify the derivation we shall neglect from the beginning any terms that are clearly of quadratic order or above when quantities are decomposed into background plus perturbation parts.

Contracting \eqref{B3}, taking its symmetric part and removing its trace gives
\begin{align}
2 \tho \left( \Phis_{ij}- { \textstyle \frac{\Phi }{d-2}} \delta_{ij} \right) 
+ \eth_{k} \left( \delta_{k (i} \Psi_{j)} - \Psi_{(ij)k} - {\textstyle \frac{2 \Psi_k }{d-2}} \delta_{ij} \right) =&
\rho_{(kl)} \left(\Phi_{ikjl} + {\textstyle \frac{2 \Phis_{kl}}{d-2}} \delta_{ij} \right) 
-\rho \left(\Phi_{ij} - {\textstyle \frac{\Phi}{d-2}} \delta_{ij} \right) \nn \\
&-\Phi \left(\rho_{ij} - {\textstyle \frac{\rho}{d-2}} \delta_{ij} \right) - { \textstyle\frac{d-4}{d-2}} \rho' \Omega_{ij}. \label{decoup:1}
\end{align}
Symmetrising over the $ij$ indices in \eqref{B5} and removing its trace gives
\begin{align}
 \eth_k \left( \Phis_{ij}- { \textstyle \frac{\Phi }{d-2}} \delta_{ij} \right)
- \eth_{l} \left( \delta_{l(i} \Phi_{j)k}- { \textstyle \frac{\Phi_{lk} }{d-2}} \delta_{ij} \right)
& +\tho' \left( \Psi_{(ij)k} + {\textstyle \frac{ \Psi_k }{d-2}} \delta_{ij} \right)
= {\textstyle{\frac{\rho}{d-2}}} \left( \Psi'_{(i}\delta_{j)k} - \Psi'_{(ij)k} - {\textstyle{\frac{2 \Psi'_{k}}{d-2}}} \delta_{ij} \right) \nn \\
& - {\textstyle{\frac{2\rho'}{d-2}}} \left(\Psi_{(ij)k} + {\textstyle{\frac{\Psi'_{k}}{d-2}}} \delta_{ij} \right) 
- {\textstyle{\frac{(d-1)\Phi}{(d-2)(d-3)}}} \left(\tau_{(i}\delta_{j)k} - {\textstyle{\frac{\tau_{k}}{d-2}}} \delta_{ij} \right). \label{decoup:2}
\end{align}
Now consider $\tho'$\eqref{decoup:1} + $\eth_k$\eqref{decoup:2}, the left hand side of which is equal to
\begin{align}
(2\tho' \tho + \eth_k \eth_k)\left( \Phis_{ij}- { \textstyle \frac{\Phi }{d-2}} \delta_{ij} \right)
&-[\tho',\eth_k] \left(\Psi_{(ij)k} + {\textstyle{\frac{\Psi'_{k}}{d-2}}} \delta_{ij} \right) \nn \\ \label{decoup:lhs}
& - \eth_{k} \left( \eth_{(i} \Phi_{j)k}- { \textstyle \frac{\eth_{l}\Phi_{lk} }{d-2}} \delta_{ij} \right)
+ \tho' \left( \eth_{(i}\Psi_{j)} - {\textstyle \frac{ \eth_k \Psi_k }{d-2}} \delta_{ij} \right).
\end{align}
The top line is precisely what we want since commutator equation \eqref{C2}$'$ can be used to convert the second term to an expression with only first order derivatives.  The second line can be simplified using the trace of \eqref{B5} and full contraction of \eqref{B7}, which when added together give
\be \label{b5b7}
\eth_{k}\Phi_{i k} - \tho' \Psi_{i} = \rho' \Psi_i + {\textstyle{\frac{d-1}{d-2}}} \Phi \tau_i.
\ee 
Applying $\eth_j$ to this, symmetrising over $ij$ and removing the trace gives
\begin{align}
&\eth_{k} \left( \eth_{(i} \Phi_{j)k}- { \textstyle \frac{\eth_{l}\Phi_{lk} }{d-2}} \delta_{ij} \right) 
- \tho' \left( \eth_{(i}\Psi_{j)} - {\textstyle \frac{ \eth_k \Psi_k }{d-2}} \delta_{ij} \right) \nn \\
&+[\eth_{k}, \eth_{l}]\left( \delta_{k(i} \Phi_{j)l}- { \textstyle \frac{\Phi_{kl} }{d-2}} \delta_{ij} \right)
+[\tho', \eth_{k}]\left(\delta_{k(i}\Psi_{j)} - {\textstyle \frac{ \Psi_k }{d-2}} \delta_{ij} \right) \nn \\
& \hspace{40mm} = \eth_{k} \left[\rho' \left(\delta_{k(i}\Psi_{j)} - {\textstyle \frac{ \Psi_k }{d-2}} \delta_{ij} \right) \right]
+ {\textstyle{\frac{d-1}{d-2}}} \eth_{k} \left[\Phi \left(\delta_{k(i}\tau_{j)} - {\textstyle \frac{ \tau_k }{d-2}} \delta_{ij} \right) \right].
\end{align}
Commutator equations \eqref{C3} and \eqref{C2}$'$ can be used to rewrite the second line in terms of first order derivative terms.  Thus, the equation above can be used to rewrite the second line of \eqref{decoup:lhs} in terms of first order derivative terms.  To summarise, we have
\be
(2\tho' \tho + \eth_k \eth_k)\left( \Phis_{ij}- { \textstyle \frac{\Phi }{d-2}} \delta_{ij} \right) = ... \, ,
\ee
where the right hand side of the equation above involves only first order derivatives.

Terms of the form $\eth_k \Psi_{(ij)k}$ and $\eth_k \Psi'_{(ij)k}$ can be removed by using equation \eqref{decoup:1}, while a term of the form $\tho' \Omega_{ij}$ can be removed by using the symmetrisation of \eqref{B2}.  

The remaining terms can be simplified by rewriting them such that one has a factor that vanishes on the background. Since, we are neglecting terms of order quadratic or above, this means the other factor takes its background value.  For example, a term of the form
\begin{align}
\rho_{(kl)} \tho' \left(\Phi_{ikjl} + {\textstyle \frac{2 \Phis_{kl}}{d-2}} \delta_{ij} \right) =&
\left( \rho_{(kl)} - {\textstyle \frac{\rho}{d-2}} \delta_{kl} \right) \tho' \left(\Phi_{ikjl} 
+ {\textstyle \frac{2 \Phis_{kl}}{d-2}} \delta_{ij} \right) 
+ {\textstyle \frac{\rho}{d-2}} \delta_{kl} \tho' \left(\Phi_{ikjl} + {\textstyle \frac{2 \Phis_{kl}}{d-2}} \delta_{ij} \right) \nn \\
=&  {\textstyle \frac{2}{(d-2)(d-3)}} \left( \rho_{(ij)} - {\textstyle \frac{\rho}{d-2}} \delta_{ij} \right) \tho' \Phi - {\textstyle \frac{2\rho}{d-2}} \tho' \left(\Phis_{ij} - {\textstyle \frac{\Phi}{d-2}} \delta_{ij} \right) \nn \\
=& - {\textstyle \frac{2(d-1)}{(d-2)^2(d-3)}} \rho' \Phi \left( \rho_{(ij)} - {\textstyle \frac{\rho}{d-2}} \delta_{ij} \right) - {\textstyle \frac{2\rho}{d-2}} \tho' \left(\Phis_{ij} - {\textstyle \frac{\Phi}{d-2}} \delta_{ij} \right),
\end{align}
where in the second equality, we have used the fact that $\rho_{(kl)} - {\textstyle \frac{\rho}{d-2}} \delta_{kl}$ vanishes on the background and since we are only considering terms that are of linear order in the perturbation expansion, we can take the background value of the factor multiplying this term.  In the final line we use  Bianchi equation \eqref{B2}$'$ evaluated on the background to simplify $\tho' \Phi$.  Using this trick to simplify all such terms and, also, using \eqref{NP4} to eliminate terms of the form $\eth_i \tau_{j}$, we find that the equation simplifies significantly
\begin{align}
\left(2\tho'\tho+ \eth_k \eth_k + {\textstyle \frac{d+2}{d-2}} (\rho' \tho + \rho \thop) + {\textstyle \frac{2(d-1)}{(d-2)^2}} \rho \rho' - {\textstyle \frac{2(d-1)(d-4)}{(d-2)(d-3)}} \Phi \right)& \left(\Phis_{ij} - {\textstyle \frac{\Phi}{d-2}} \delta_{ij} \right) \nn \\ \label{phitdecoup}
& + {\textstyle \frac{(d-4)}{(d-2)^2}} \left( \rho'^2 \Om_{ij}+ \rho^2 \Om'_{ij} \right) =0.
\end{align}
Of course, we would like to derive an equation satisfied by $\rho^2 \left(\Phis_{ij} - {\textstyle \frac{\Phi}{d-2}} \delta_{ij} \right)$.  Letting ${\Phis}^t_{ij}=\left(\Phis_{ij} - {\textstyle \frac{\Phi}{d-2}} \delta_{ij} \right)$,
\begin{gather*}
 \rho^2 \tho' \tho {\Phis}^t_{ij} = \thop\left( \rho^2 \tho {\Phis}^t_{ij}\right) - (\thop \rho^2) \tho {\Phis}^t_{ij}
= \thop \tho  \left( \rho^2 {\Phis}^t_{ij}\right) - (\thop \rho^2) \tho {\Phis}^t_{ij} - (\tho \rho^2) \thop {\Phis}^t_{ij} - (\thop \tho \rho^2) {\Phis}^t_{ij}, \\
\rho^2 \tho {\Phis}^t_{ij} = \tho \left( \rho^2 {\Phis}^t_{ij}\right) - (\tho \rho^2) {\Phis}^t_{ij}, \\
\rho^2 \thop {\Phis}^t_{ij} = \thop \left( \rho^2 {\Phis}^t_{ij}\right) - (\thop \rho^2) {\Phis}^t_{ij}.
\end{gather*}
Equations \eqref{NP1} and \eqref{NP2} evaluated on the background give
\be
\tho \rho = - {\textstyle \frac{\rho^2}{d-2}}, \quad \thop \rho = - {\textstyle \frac{\rho \rho'}{d-2}} - \Phi.
\ee
Thus, equation \eqref{phitdecoup} is equivalent to equation \eqref{rphitdecoup}
\begin{align}
\left(2\tho'\tho+ \eth_k \eth_k + {\textstyle \frac{d+6}{d-2}} (\rho' \tho + \rho \thop) + {\textstyle \frac{4 \Phi}{\rho}} \tho + {\textstyle \frac{2(3d+5)}{(d-2)^2}} \rho \rho' - {\textstyle \frac{4(3d-8)}{(d-2)(d-3)}} \Phi \right) & \left[ \rho^2 \left(\Phis_{ij} - {\textstyle \frac{\Phi}{d-2}} \delta_{ij} \right) \right]  \nn \\ \nn
& + {\textstyle \frac{(d-4)}{(d-2)^2}} \rho^2 \left( \rho'^2 \Om_{ij}+ \rho^2 \Om'_{ij} \right) =0.
\end{align}

\section{Adjoints of GHP covariant derivatives}

In this appendix, we derive the adjoints of the GHP covariant derivatives.  First, consider the adjoint of operator $\tho$.  Let $\eta_{i_1 \ldots i_{s}}$ and $\zeta_{i_1 \ldots i_{s}}$ be GHP scalars of spin $s$ and boost weights $b$ and $-(b+1)$, respectively and consider the inner product
\begin{align*}
(\zeta_{i_1 \ldots i_{s}}, \tho \, \eta_{i_1 \ldots i_{s}})&=\left( \zeta_{i_1 \ldots i_{s}} , \ell \cdot \partial \, \eta_{i_1 i_2...i_s} - b L_{10} \eta_{i_1 i_2...i_s} 
                + {\textstyle{\sum_{r=1}^s \M{k}_{i_r 0}}} \, \eta_{i_1...i_{r-1} k i_{r+1}...i_s}  \right) \\
&=\left(- b L_{10} \zeta_{i_1 i_2...i_s} 
   - {\textstyle{\sum_{r=1}^s \M{k}_{i_r 0}}} \, \zeta_{i_1...i_{r-1} k i_{r+1}...i_s}, \ell \cdot \partial \, \eta_{i_1 i_2...i_s} \right) \\
&=\left(- [(\ell \cdot \partial + \nabla \cdot \ell) \, \zeta_{i_1 i_2...i_s} + b L_{10} \zeta_{i_1 i_2...i_s} 
                + {\textstyle{\sum_{r=1}^s \M{k}_{i_r 0}}} \, \zeta_{i_1...i_{r-1} k i_{r+1}...i_s}], \eta_{i_1 i_2...i_s} \right) \\
&=\left(- [(\ell \cdot \partial + \rho) \, \zeta_{i_1 i_2...i_s} + (b+1) L_{10} \zeta_{i_1 i_2...i_s} 
                + {\textstyle{\sum_{r=1}^s \M{k}_{i_r 0}}} \, \zeta_{i_1...i_{r-1} k i_{r+1}...i_s}], \eta_{i_1 i_2...i_s} \right) \\
&=\left(- [\tho + \rho] \zeta_{i_1 i_2...i_s}, \eta_{i_1 i_2...i_s} \right),
\end{align*}
where the first equality uses the definition of operator $\tho$ given in eq. \eqref{ghpdertho}, the second equality uses the property that
\be
\M{i}_{j \mu} + \M{j}_{i \mu} = 0
\ee
and the third inequality is obtained using integration by parts and ignoring divergence terms, since operator adjoints are defined up to such terms.  The penultimate equality uses the geodesity of $\ell$ to deduce that
\be
\nabla \cdot \ell = L_{10} + \rho
\ee
and the final equality uses the definition of operator $\tho$ given in eq. \eqref{ghpdertho}. Hence,
\be \label{thoad}
\tho^\dag=-(\tho +\rho).
\ee

Taking the prime of this equation gives the adjoint of $\thop$
\be \label{thopad}
\thop^\dag=-(\thop +\rho').
\ee

Now, consider the inner product of $\eth_{i_{1}} \eta_{i_2...i_{s+1}}$ with $\xi_{i_1...i_{s+1}}$, a GHP scalar with boost weight $-b$ and spin $s+1$
\begin{align*}
(\xi_{i_1...i_{s+1}}, \eth_{i_{1}} \eta_{i_2...i_{s+1}})&=\left( \xi_{i_1...i_{s+1}}, [\mb {i_{1}} \cdot \partial - b L_{1i_{1}}] \eta_{i_2...i_{s+1}} 
                                     + {\textstyle{\sum_{r=2}^{s+1} \M{k}_{i_r i_1}}} \eta_{i_2...i_{r-1} k i_{r+1}...i_{s+1}} \right) \\
&=\left(- [b L_{1i_{1}} - \M{i_1}_{k k} ] \xi_{i_1...i_{s+1}} 
                                     - {\textstyle{\sum_{r=1}^{s+1} \M{k}_{i_r i_1}}} \xi_{i_1..i_{r-1} k i_{r+1}...i_{s+1}}, \mb {i_{1}} \cdot \partial \eta_{i_2...i_{s+1}} \right) \\
&=\left(- [\mb{i_1} \cdot \partial + \nabla \cdot \mb{i_1} + b L_{1i_{1}} - \M{i_1}_{k k} ] \, \xi_{i_1...i_{s+1}} \right. \\
                            &  \left. \hspace{60mm}  - {\textstyle{\sum_{r=1}^{s+1} \M{k}_{i_r i_1}}} \xi_{i_1..i_{r-1} k i_{r+1}...i_{s+1}}, \eta_{i_2...i_{s+1}} \right) \\
&=\left(- [\eth_{i_1} + \nabla \cdot \mb{i_1} - \M{i_1}_{k k}] \xi_{i_1...i_{s+1}}, \eta_{i_2...i_{s+1}} \right) \\
&=\left(- [\eth_{i_1} - \tau_{i_1} - \tau'_{i_1}] \xi_{i_1...i_{s+1}}, \eta_{i_2...i_{s+1}} \right),
\end{align*}
where the first equality uses the definition of operator $\eth$ given in eq. \eqref{ghpderm}, the second equality uses the property that
\be
\M{i}_{j \mu} + \M{j}_{i \mu} = 0
\ee
and the third inequality is obtained using integration by parts.  The penultimate equality uses the definition of operator $\eth$ and the final equality uses the fact that
\be
\nabla \cdot \mb{i} =  \M{i}_{k k} - \tau_i - \tau'_i.
\ee
Thus,
\be \label{ethad}
{\eth_i}^\dag =-\eth_i + \tau_i + \tau'_i.
\ee

\end{document}